\documentclass{PoS}
\usepackage{amsmath}

\newcommand{\be}{\begin{equation}}

\newcommand{\ee}{\end{equation}}
\newcommand{\bea}{\begin{eqnarray}}
\newcommand{\eea}{\end{eqnarray}}

\newcommand{\nn}{\nonumber}
\newcommand{\de}{\partial}
\def\eps{\epsilon}

\def\Wt{\tilde{ {W}}}

\def\Yt{\tilde{ {Y}}}

\def\gt{\tilde g}

\def\yyt{{\tilde{\cal Y}}_\mu}

\title{Fermion Delocalization in Higgsless Models}

\ShortTitle{Fermion Delocalization in Higgsless Models}

\author{\speaker{Stefania De Curtis}\thanks{In collaboration with R.~Casalbuoni, D.~Dominici and D.~Dolce.}\\
        INFN Firenze, Italy\\
        E-mail: \email{decurtis@fi.infn.it}}

\abstract{In the linear moose framework, which naturally emerges in
deconstruction models, we discuss the effect of direct couplings
between the left-handed fermions living on the boundary of the chain
and the gauge fields in the internal sites. This is realized by
means of a product of nonlinear sigma-model scalar fields which, in
the continuum limit, is equivalent to a Wilson line. The effect of
these new nonlocal couplings is a contribution to the S parameter
which can be of opposite sign with respect to the one coming from
the gauge fields along the string. Therefore, with some fine-tuning,
it is possible to satisfy the constraints from the electro-weak data
without spoiling the perturbative unitarity limit, which, in these
models is generally postponed with respect to the Higgsless Standard Model
one.}

\FullConference{International Europhysics Conference on High Energy Physics\\
         July 21st - 27th 2005\\
         Lisboa, Portugal}

\begin{document}

\section{Higgsless electro-weak symmetry breaking from moose models}

\begin{figure}[h]
\begin{center}
  \includegraphics[width=.6\textwidth]{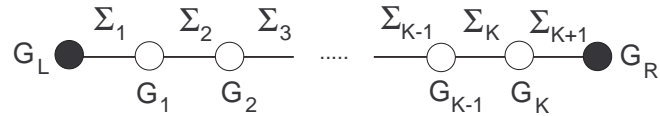}
  \caption{The simplest moose diagram for the Higgsless breaking of the EW symmetry.
  \label{fig:1}}
  \end{center}
\end{figure}

Higgsless models may represent an alternative to the standard
electro-weak (EW) symmetry breaking mechanism. In the past few
years,  after the blooming of extra-dimensions, they have received a
renewal of interest.
 Higgsless models, in their "modern" version,  are
formulated as gauge theories in a five dimensional space and, after
decompactification, describe a tower of Kaluza Klein (KK)
excitations of the standard EW gauge bosons \cite{ref}. One of the
interesting features of these schemes is the possibility to delay
the unitarity violation scale via the exchange of massive KK modes
\cite{ref}. However,  it is generally difficult to reconcile a
delayed unitarity with the EW constraints.
 For instance in
 the framework of  models with only ordinary fermions, it is possible to get a small or
zero $S$ parameter \cite{moose}, at the expenses of having a
unitarity bound as in the Standard Model (SM) without the Higgs,
that is of the order of 1 TeV. A recent solution to the problem
which does not spoil the unitarity requirement at low scales, has
been found by delocalizing the fermions in five dimensional theories
\cite{Cacciapaglia:2004rb,Foadi:2004ps}. We will investigate this
possibility in the context of  deconstructed gauge theories which
come out when the extra dimension is discretized \cite{decon}.
Through discretization of the fifth dimension we get a finite set
of four-dimensional gauge theories each of them acting at a
particular lattice site. In this construction, any connection field
along the fifth dimension, $A_5$, goes naturally into the link
variables $\Sigma_i= e^{-ia A_5^{i-1}}$ realizing the parallel transport
between two lattice sites (here $a$ is the lattice spacing).
 They satisfy the condition $\Sigma\Sigma^\dagger =1$ and  can be identified with  chiral
fields.
   In this way
the discretized version of the original 5-dimensional gauge theory
is substituted by a collection of four-dimensional gauge
theories with gauge interacting chiral fields $\Sigma_i$,  synthetically described by a moose
diagram (an example is given in Fig. \ref{fig:1}).

Here we will  consider the simplest  linear moose model for the Higgsless breaking of the EW symmetry
and we will delocalize fermions by introducing direct couplings between ordinary left-handed fermions and the gauge
vector bosons  along the moose string \cite{ferm}.

Let us briefly review the linear moose model based on the $SU(2)$ symmetry \cite{moose,ferm}.
  We consider
$K+1$ non linear $\sigma$-model scalar fields $\Sigma_i$, ${i=1,\cdots ,K+1}$,
$K$ gauge groups, $G_i$, ${i=1,\cdots ,K}$
 and a global symmetry $G_L\otimes
G_R$ as shown in Fig. \ref{fig:1}.
A minimal model of EW symmetry breaking
 is obtained by choosing $G_i=SU(2)$,
$G_L\otimes G_R=SU(2)_L\otimes SU(2)_R$. The SM
gauge group
  $SU(2)_L\times U(1)_Y$ is obtained by gauging
a subgroup of $G_L\otimes G_R$. The $\Sigma_i$ fields can be
parameterized as $\Sigma_i=\exp{[i/(2f_i)\vec \pi_i\cdot \vec
\tau}]$ where $\vec \tau$ are the Pauli matrices and $f_i$ are $K+1$
constants that we will call link couplings. The lagrangian of the
linear moose model is given by \be {\cal
L}=\sum_{i=1}^{K+1}f_i^2{\rm Tr}[D_\mu\Sigma_i^\dagger
D^\mu\Sigma_i]-\frac 1 2\sum_{i=1}^K{\rm Tr}[(F_{\mu\nu}^i)^2]
-\frac 1 2{\rm Tr}[(F_{\mu\nu}(\Wt))^2 -\frac 1 2{\rm
Tr}[(F_{\mu\nu}(\Yt))^2] , \label{lagrangian:l} \ee with the
covariant derivatives  defined as follows:
$D_\mu\Sigma_1=\de_\mu\Sigma_1-i\gt \Wt_\mu\Sigma_1+i\Sigma_1 g_1
V_\mu^1$;
$~D_\mu\Sigma_i=\de_\mu\Sigma_i-ig_{i-1}V_\mu^{i-1}\Sigma_i+i\Sigma_i
g_i V_\mu^i~~(i=2,\cdots,K)$;
$~D_\mu\Sigma_{K+1}=\de_\mu\Sigma_{K+1}-ig_{K}V_\mu^{K}\Sigma_{K+1}+i
\tilde g'\Sigma_{K+1}\Yt_\mu$, where $V_\mu^i=V_\mu^{ia}\tau^a/2$
and $g_i$ are the gauge fields and gauge coupling constants
associated to the groups $G_i$, $i=1,\cdots ,K$, and
$\Wt_\mu=\Wt_\mu^{a}\tau^a/2$, $\Yt_\mu=\yyt\tau^3/2$ are the gauge
fields associated to $SU(2)_L$ and $U(1)_Y$ respectively. Notice
that, in the unitary gauge, all the $\Sigma_i$ fields are eaten up
by the gauge bosons which acquire mass, except for the photon
corresponding to the unbroken $U(1)_{em}$. By identifying the lowest
mass eigenvalue in the charged sector at $O(\gt^2/g_i^2)$ with
$M_W$, we get a relation between the EW scale v ($\approx 250~GeV$)
and the link couplings of the chain: \be \frac{4}{{\rm
v}^2}\equiv\frac 1 {f^2}=\sum_{i=1}^{K+1}\frac 1 {f_i^2}.\nn
\label{eq:20}\ee Concerning fermions, we will consider only the
standard model ones, that is:
 left-handed fermions  $\psi_L$ as  $SU(2)_L$
doublets and  singlet right-handed fermions
$\psi_R$ coupled to the SM gauge fields through the
groups $SU(2)_L$ and $U(1)_Y$ at the ends of the chain.

\section{Constraints from perturbative unitarity and EW tests}

The worst high-energy behavior of the moose models arises from the
scattering of longitudinal vector bosons. To simplify the
calculation we will make use of the equivalence theorem, that is of
the possibility of evaluating this amplitude in terms of the
scattering amplitude of the corresponding Goldstone bosons. However this theorem holds in the
approximation where the energy of the process is much higher than the
mass of the vector bosons.  Let us evaluate the amplitude for the SM $W$ and $Z$ at
energies $M_{W/Z}\ll E\ll M_{V_i}$. The unitary
gauge for the $V_i$ bosons is given by the choice $\Sigma_i=\exp[i
f\vec\pi\cdot\vec\tau/(2f_i^2)]$ with $f$ given in eq.
(\ref{eq:20}) and $\vec\pi$ the GB's giving mass to $W$ and $Z$. The resulting four-pion amplitude is \be
A_{\pi^+\pi^-\to\pi^+\pi^-}=-\frac{f^4 u}4\sum_{i=1}^{K+1}\frac 1
{f_i^6}+\frac {f^4}4\sum_{i,j=1}^K
L_{ij}\left((u-t)(s-M_2)_{ij}^{-1}+(u-s)(t-M_2)_{ij}^{-1}\right),\label{uni}\ee
with $(M_2)_{ij}$ the square mass matrix for the gauge fields, and
$ L_{ij}=g_ig_j(f_i^{-2}+f_{i+1}^{-2})(f_j^{-2}+f_{j+1}^{-2})$.
In the high-energy limit, where
we can neglect the second term  in eq. (\ref{uni}),  the amplitude has a minimum for all the $f_i$'s
being equal to a common value $f_c$. As a consequence, the scale at which unitarity is violated
by this single channel contribution  is delayed by a factor $(K+1)$ with respect to the one in the SM without the Higgs:
$\Lambda_{moose}=(K+1)\Lambda_{HSM}$.

However the moose model has many other longitudinal vector bosons
with bad behaving scattering amplitudes.
For  energies much higher than all the
masses of the vector bosons, we can determine the unitarity
bounds by considering  the eigenchannel amplitudes corresponding
to all the possible four-longitudinal vector bosons. Since the
unitary gauge for all the vector bosons is  given by $\Sigma_i=\exp[i
\vec\pi\cdot\vec\tau/(2f_i)]$, the amplitudes are already diagonal, and
the high-energy result is simply $
A_{\pi_i\pi_i\to\pi_i\pi_i}\to -{u}/(4f_i^2)$. We see that, also in this case, the best
unitarity limit is for all the link couplings being equal: $f_i=f_c$.
Then: $\Lambda^{TOT}_{moose}=\sqrt{K+1}~\Lambda_{HSM}$ (for similar results see ref. [23] in \cite{ferm}).
 However, in order our approximation to be
correct, we have to require $M_{V_i}^{max}\ll \Lambda^{TOT}_{moose}$. By using the explicit expression for the
highest mass eigenvalue, in the case of equal couplings $g_i=g_c$, we get an upper bound
 $g_c <5$. As we will see, this choice gives  unacceptable large EW correction.

In this class of models all the corrections from new physics are "oblique" since they arise from
mixing of the SM vector bosons with the moose vector fields (we are assuming the standard couplings for the
fermions to $SU(2)_L\otimes U(1)$).
 As well known, the oblique
corrections are completely captured by the parameters $S$, $T$ and
$U$ or, equivalently by the
parameters $\epsilon_i$, $i=1,2,3$. For the linear moose, the
existence of the custodial symmetry $SU(2)_V$  ensures that
$\epsilon_1\approx \epsilon_2\approx 0$. On the contrary,
 the new physics contribution to the EW parameter
$\epsilon_3$  is sizeable and positive \cite{moose}:
$ \epsilon_3=(\gt^2/g_i^2)\sum_{i=1}^K(1-y_i)y_i$,
where  $ y_i=\sum_{j=1}^i f^2/f_j^2$.
   Since
$0\le y_i\le 1$ it follows $\epsilon_3\ge 0$ (see also
\cite{barb,hirn,georgi}). As an example,
let us take equal couplings along the chain: $f_i=f_c$,  $g_i=g_c$. Then  $\epsilon_3=
\gt^2 ~K(K+2)/(6~g_c^2(K+1))$, which grows with the number of sites of the moose. If we want to be
compatible with the experimental data we need to get
$\epsilon_3\approx 10^{-3}$. Already for  $K=1$ this would require $g_c\ge
15.8 \gt$, implying a strong interacting gauge theory in the moose
sector and unitarity violation. Notice also that, insisting on a weak gauge theory would
imply $g_c$ of the order of $\gt$, then the natural value of $\epsilon_3$ would be of the order
$10^{-1}-10^{-2}$, incompatible with the experimental data.

\section{Effects of fermion delocalization}

A way to reconcile perturbative unitarity requirements with the EW bounds is to
allow delocalized couplings of the SM fermions to the moose gauge
fields and some amount of fine tuning \cite{ferm}. In fact, by genaralizing the procedure in
\cite{cas}, the SM fermions can
be coupled to any of the gauge fields  at the lattice sites
by means of a Wilson line.
 Define $
\chi_L^i=\Sigma_i^\dagger\Sigma_{i-1}^\dagger\cdots\Sigma_1^\dagger\psi_L$, for $i=1,\cdots,K$.
Since under a gauge transformation, $\chi_L^i\to U_i\chi_L^i$, with
$U_i\in G_i$, at each site we can introduce a gauge
invariant coupling given by
\be
b_i\bar\chi_L^i\gamma^\mu\left(\de_\mu+ig_i V_\mu^i+\frac i 2
{\tilde g}'(B-L)\Yt_\mu\right)\chi_L^i,\ee
where $B(L)$ is the barion(lepton) number and $b_i$ are dimensionless parameters.
The new fermion interactions give extra non-oblique contributions to the EW parameters.
These are calculated in \cite{ferm} by decoupling the $V_\mu^i$ fields and evaluating the corrections to the relevant
physical quantities.
 To the first order in $b_i$ and to $O(\gt^2/g_i^2)$, the $\epsilon_i$ parameters are modified as follows:
 \be \epsilon_1\approx 0,~~~\epsilon_2\approx
0,~~~\epsilon_3\approx\sum_{i=1}^K
y_i\left(\frac{g^2}{g_i^2}(1-y_i)-b_i\right).\label{tuning}\ee
This final expression suggests that the introduction of
the $b_i$ direct fermion couplings to $V_i$ can compensate for the
contribution of the tower of gauge vectors to $\eps_3$. This would
reconcile  the Higgsless model with the EW precision
measurements by fine-tuning the direct fermion couplings.

\begin{figure}[ht]
\begin{center}
\includegraphics[width=.30\textwidth]{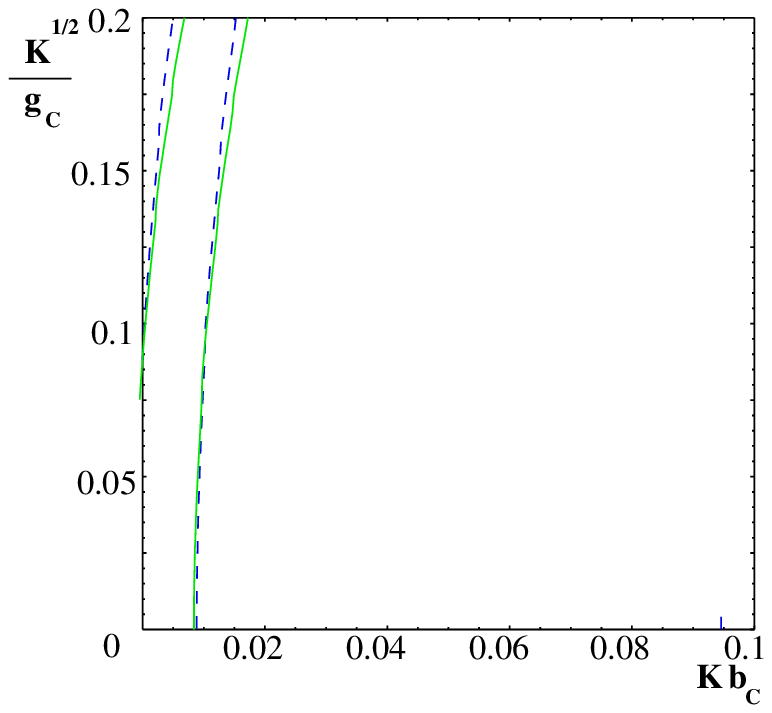}
  \includegraphics[width=.30\textwidth]{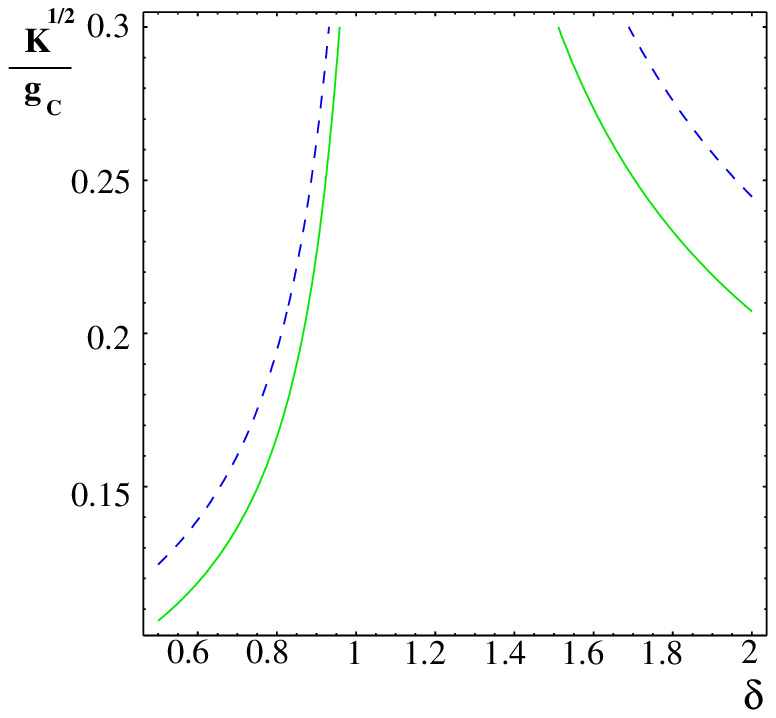}
   \caption{
The 95\% CL bounds on the plane $(Kb_c,\sqrt{K}/g_c)$-left panel, $(\delta,\sqrt{K}/g_c)$-right panel,
 from the experimental value
of $\eps_3$ for $K=1$ (solid green lines), $K=10$
 (dash blue lines). The allowed regions are between the corresponding lines.
\label{fig:2}}
  \end{center}
\end{figure}

In the simplest model  with all $f_i=const=f_c$, $g_i=const=g_c$
and $b_i=const=b_c$, as shown in the left panel of Fig. \ref{fig:2},
the experimental bounds from the $\eps_3$ parameter
can be satisfied by fine-tuning the direct fermion coupling $b_c$
along a strip in the plane $(Kb_c,\sqrt{K}/g_c)$ (we have chosen these
parameters due to the scaling properties of $g_c$ and $b_c$ with $K$, see ref. \cite{ferm} for details).

The expression for  $\eps_3$ given in eq. (\ref{tuning})
suggests also the possibility of a site-by-site cancellation, provided by:
$ b_i=\delta (g^2/g_i^2)(1-y_i)$.
This choice, for small $b_i$,
gives $\eps_3\approx 0$ for $\delta=1$.
Assuming again  $f_i=f_c$, $g_i=g_c$, the allowed
region in the space $(\delta,\sqrt{K}/g_c)$ is given on the right panel of Fig.
\ref{fig:2}.

In conclusion, by fine tuning every direct fermion
coupling at each site  to compensate the corresponding contribution
to $\eps_3$ from the moose gauge bosons (see also \cite{chiv}),
it is possible to satisfy the EW
constraints and improve the unitarity bound of the Higgsless SM at
the same time.

\end{document}